\documentclass[12pt, a4paper]{article}
\usepackage{geometry}
\usepackage[utf8]{inputenc}  
\usepackage[T1]{fontenc}     
\usepackage{lmodern}         
\usepackage{amsmath,amssymb} 

\usepackage{times,color}
\usepackage{amsmath}
\usepackage{subfigure, epsfig, graphicx, epstopdf}
\usepackage[colorlinks]{hyperref}
\hypersetup{linkcolor = {blue},	citecolor = {blue}, urlcolor = {blue}}
\usepackage{cite}
\usepackage{amssymb}
\usepackage[amssymb]{SIunits}

\newenvironment{sciabstract}{%
\begin{quote} \bf}
{\end{quote}}

\newcounter{lastnote}

\begin{document}

\title{Spatiotemporal Topological Combs for Robust High-Dimensional Information Transmission}
\author{Dawei Liu$^{1 \dag}$,  Daijun Luo$^{1 \dag}$,  Huiming Wang$^{3 \dag}$, Xingyuan Zhang$^{2}$,\\
Zhirong Tao$^{4}$, Dana JiaShaner$^{1}$, Zhensheng Tao$^{5}$\\
Qian Cao$^{1}$, Xiaoshi Zhang$^{6}$, Guangyu Fan$^{1\ast}$, Qiwen Zhan$^{1, 2, 7}$ \\
\normalsize{$^1$School of Optical-Electrical and Computer Engineering, }\\
\normalsize{University of Shanghai for Science and Technology, 200093, Shanghai, China} \\
\normalsize{$^2$Zhejiang Key Laboratory of 3D Micro/Nano Fabrication and Characterization,}\\ \normalsize{Department of Electronic and Information Engineering, School of Engineering,}\\
\normalsize{Westlake University, Hangzhou, Zhejiang 310030, China}\\
\normalsize{$^3$Institute of Photonics, TU Wien, Gusshausstrasse 27/387, Vienna, Austria}\\
\normalsize{$^4$Karlsruhe Institute of Technology, Karlsruhe, Germany}\\
\normalsize{$^5$State Key Laboratory of Surface Physics, }\\
\normalsize{Key Laboratory of Micro and Nano Photonic Structures (MOE), }\\
\normalsize{Department of Physics, Fudan University, Shanghai 200433, China}\\
\normalsize{$^6$Yunnan University, Kunming, Yunnan, 650500, China }\\
\normalsize{$^7$International Institute for Sustainability with Knotted Chiral Meta Matter (WPI-SKCM2),}\\
\normalsize{Hiroshima University, Higashihiroshima, Hiroshima, 739-8526, Japan}\\
\normalsize{$^\dag$These authors contributed equally to this work.}\\
\normalsize{$^\ast$E-mail: gfan@usst.edu.cn}\\
}

\date{\today}

\baselineskip24pt

\maketitle

\begin{sciabstract}
Sculpting light across its independent degrees of freedom—from orbital angular momentum to the discrete wavelengths of optical frequency combs—has unlocked vast communication bandwidth by enabling massively parallel information channels. However, the Shannon–Hartley theorem sets a hard limit by tying channel capacity to the trade-off between SNR and rate, a central challenge in communication. Inspired by lock-in amplification in electronics, we encode data on THz optical burst carriers so the signal resides beyond the conventional noise band, yielding exceptional robustness. By leveraging a programmable all-degree-of-freedom (All-DoF) modulator, we generate a spatiotemporal topological comb (ST-Comb) that structures light into a vast, high-entropy state space for high-dimensional information encoding. Crucially, we find that the associated topological winding number is preserved under diverse perturbations, ensuring stable information encoding and retrieval. This paradigm illustrates how structured light can simultaneously expand channel dimensionality and maintain robustness, charting a pathway to chip-scale, reconfigurable photonic platforms for the PHz era, while also opening previously inaccessible regimes of light–matter interaction.
\end{sciabstract}

\section*{Introduction}

\noindent The explosive growth of AI-driven data is rapidly outpacing the capacity of existing optical links, creating an urgent need for ultra-high-capacity, low-latency, and high-fidelity communication\cite{fu2024optical}. To address these challenges, light’s intrinsic multiplexing capacity offers tremendous potential, with two emerging directions attracting particular attention: orbital angular momentum (OAM) multiplexing \cite{wang2012terabit,bozinovic2013terabit} and wavelength-division multiplexing (WDM) \cite{del2007optical,shu2022microcomb,armani2003ultra}. Owing to its unbounded set of orthogonal modes, OAM significantly enhances the parallelism of information transmission. Simultaneously, WDM scales capacity by parallelizing channels, with optical frequency combs supplying evenly spaced carriers for dense implementations that markedly boost data rates and total capacity\cite{marin2017microresonator,spencer2018optical,suh2016microresonator}. The advancement of micro/nano fabrication technology has facilitated the emergence of on-chip optical frequency combs, which enhance performance, reduce system power consumption and size, and provide a flexible, cost-effective pathway to next-generation optical communication systems\cite{gaeta2019photonic,sun2023applications,chang2022integrated}.

Meanwhile, the Shannon–Hartley theorem elucidates the fundamental limits of information transmission, indicating a trade-off between the transmission rate and the signal-to-noise ratio (SNR) under fixed bandwidth conditions\cite{jorgensen2022petabit,chen2024quantum}. Despite significant progress in reducing phase noise and enhancing amplitude stability in Kerr frequency combs\cite{marin2017microresonator,herr2012universal,zhang2024advances}, minimizing noise in optical channels and further improving the SNR remain major challenges. In electronics, lock-in amplification has long been a key technique for effectively extracting weak signals from noise\cite{michels1941pentode}, and its fundamental principles can be generalized to optical systems\cite{kotler2011single,chen2024quantum}. Building on this concept, we introduce an optical lock-in encoding and decoding mechanism. By encoding spatiotemporal topological information onto high-frequency carriers (ultrafast pulse arrays ranging from THz–PHz), this scheme effectively suppresses noise in the GHz range typical of conventional communication systems, resulting in an exceptionally high SNR. Leveraging an all-degree-of-freedom (All-DoF) modulator, we realize a programmable spatiotemporal topological comb (ST-Comb), which enables high-dimensional information encoding. We further introduce the tunable spatiotemporal carrier-envelope phase (ST-CEP), extending the traditional CEP concept into the spatiotemporal domain and establishing its intrinsic link to conventional CEP, thereby providing new opportunities for controlling electromagnetic fields at the sub-cycle level\cite{baltuvska2003attosecond,sussmann2015field,goulielmakis2022high}. Furthermore, we reveal the interplay between temporal and spectral encoding dimensions, creating new implementation pathways for complex spatiotemporal information processing.

Based on these principles, we develop a multi-dimensional (de)multiplexing framework within a single wavepacket, including sub-pulse parallel multiplexing, ST-CEP multiplexing, and radial index multiplexing. These additional degrees of freedom allow precise control over the ultrafast spatiotemporal topological spectral structure. Importantly, we demonstrate that the winding number associated with the ST-Comb remains invariant under a variety of perturbations, ensuring reliable and stable information encoding. This invariance further enables a deep learning framework, trained on lock-in-decoded data, to extract high-fidelity information and perform system-level multiplexing and demultiplexing. Collectively, these results demonstrate the robustness and scalability of ST-Comb, highlighting its potential to establish a new paradigm for next-generation optical communication with unprecedented bandwidth and noise resilience.

\section*{Concepts of ST-Comb and Optical Lock-In}
\begin{figure*}
\centering
\includegraphics[width=1\columnwidth]{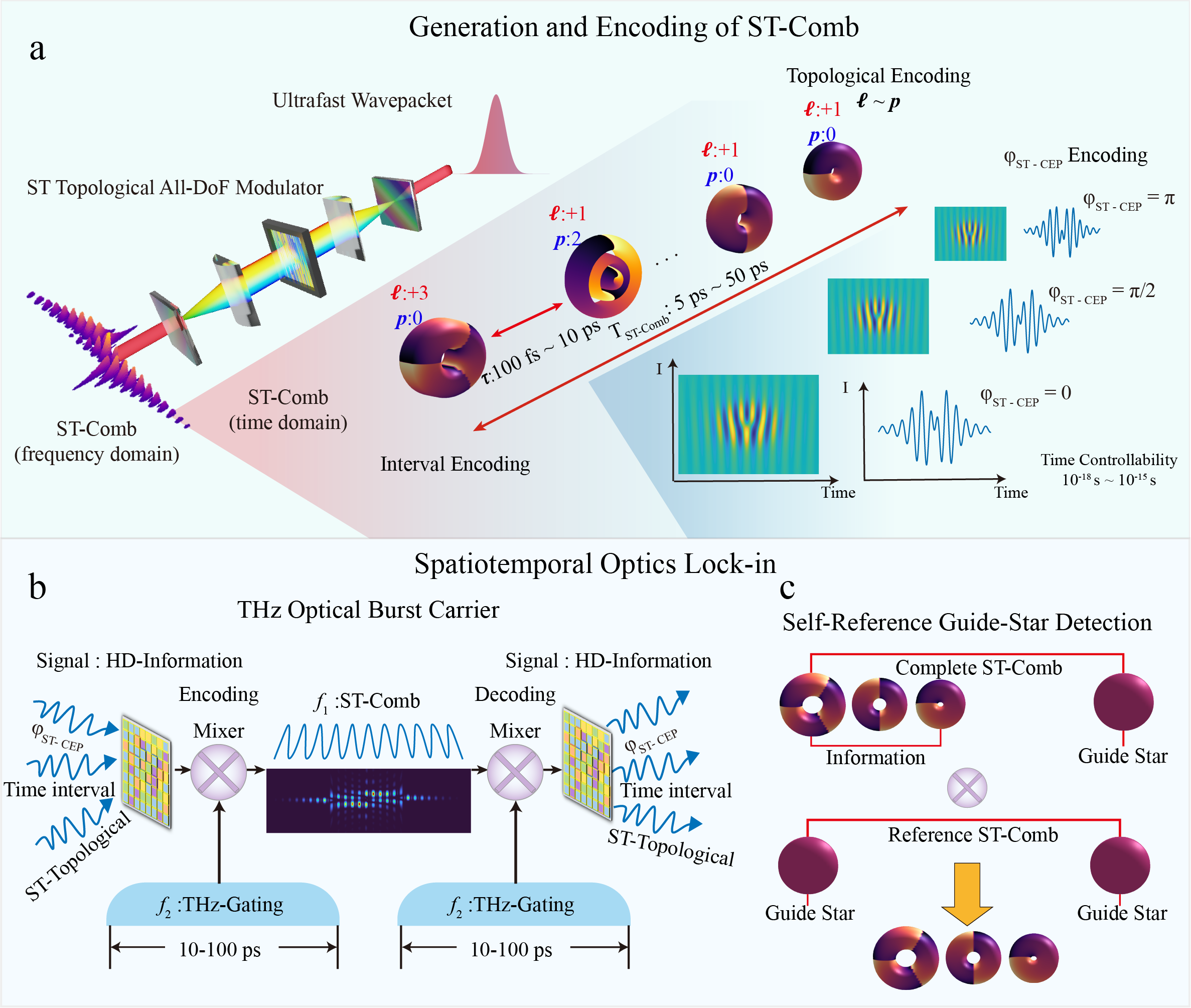}
\caption{\textbf{Generation of the ST-Comb and spatiotemporal optical lock-in.}  \textbf{a,} Formation and encoding. An ultrafast pulse is shaped by a spatiotemporal modulator into a sequence of spatiotemporal-topological sub-pulses. The fs–ps inter-pulse spacing corresponds to a THz-rate carrier. 
The topological modes (including radial index $\mathit{p}$ and azimuthal index $\ell$), temporal separation $\tau$, and relative phase difference $\boldsymbol{\varphi}_{ST-CEP}$ of each sub-pulse can all be utilized for information encoding. \textbf{b,} Spatiotemporal lock-in encoding. All spatiotemporal information is loaded onto a THz-burst carrier via the modulator, achieving spatiotemporal phase lock-in, heterodyning with an information-free Gaussian THz gating pulse retrieves the encoded spatiotemporal content. \textbf{c,} Self-reference “Guide-Star” Detection. The first sub-pulse is set as a Gaussian guide-star and temporally separated from the information-bearing sub-pulses, subtracting the two-Gaussian reference ST-Comb isolates the spatiotemporal topological information.}
 \label{Figure1}
\end{figure*}

Here we establish the concept of optical lock-in and use a spatiotemporal topological comb (ST-Comb) as a model system to verify. In the time domain, the ST-Comb manifests as a sequence of ultrafast pulses (Figure~\ref{Figure1}a). Each sub-pulse carries individually addressable topological degrees of freedom—a counterintuitive observation given that topological information is inherently encoded in the spectral domain ($k_x$–$\omega$ plane). This phenomenon arises because the spiral phase is preserved through a 2D Fourier transform during spectral-to-spatiotemporal mapping\cite{chong2020generation,wan2022toroidal}. Prior studies have shown that $4\mathit{f}$ optical shapers enable the generation of Spatiotemporal Optical Vortices (STOV), as their spatial and temporal profiles are strongly coupled—the pulse shape evolves spatially. Building on this foundation, we extend the concept to broader spatiotemporal sequences by introducing a universal framework for generating ST-Combs. Specifically, we propose a systematic approach for encoding single-pulse wave packets via a spatiotemporal all-degrees-of-freedom modulator (All-DoF modulator). This transforms an input femtosecond pulse into a programmable sequence of ultrafast pulses with tailored spatiotemporal topological degrees of freedom, thereby defining the ST-Comb. The electric field distribution of the generated comb is formulated as (Method):

\begin{equation}
E(x, y, t) = \sum_{n} C_n \, A_n(x, y; \mu_n) \exp \bigl\{ i \bigl[ \omega_0 t + \psi_{\text{ST-CEP}}(n) \bigr] \bigr\},
 \end{equation}

Here, $C_n$ denotes the peak amplitude of the n-th sub-pulse, $A_n$ the normalized amplitude distribution of the n-th sub-pulse, and $\mu_n$ characterizes the topological degree of freedom (orbital angular momentum quantum number $\ell$ or radial mode number $\mathit{p}$). $\psi_\text{ST-CEP}$ corresponds to the spatiotemporal carrier-envelope phase (ST-CEP) of the sub-pulse. By modulating $\psi_\text{ST-CEP}$, the spectral shift $\Delta \nu = {\Delta \psi}/{2\pi \tau}$ (where $\tau$ denotes the interval between sub-pulses) can be precisely controlled. This approach enables the generation of complex spatiotemporal topological structures—including both orthogonal and non-orthogonal configurations—without requiring computationally intensive reconstruction algorithms (Supplementary Materials 1).

Lock-in detection provides a powerful analogy: detecting a faint signal in overwhelming noise by shifting it to a frequency band where noise is minimal. In optics, this principle is realized by encoding information on THz–PHz carriers, which shifts the signal far beyond the conventional noise band (typically below GHz) and physically isolates it from environmental fluctuations. The information-bearing ST-Comb is then heterodyned with a co-propagating reference carrier, allowing phase-sensitive (synchronous) detection to recover the encoded spatiotemporal content with significant SNR improvement (Figure~\ref{Figure1}b). Beyond noise suppression, the ST-Comb exhibits topological protection: while spatiotemporal astigmatism during propagation may distort waveform details, the global topological charge remains invariant\cite{hancock2021mode,gui2021second}. This invariance, reminiscent of the robustness of Winding numbers in complex-field topology, ensures that local perturbations do not compromise the encoded information (Supplementary Materials 3).

\section*{Temporal and Spectral Properties of ST-CEP and ST-Comb}
\begin{figure*}
\centering
\includegraphics[width=0.80\columnwidth]{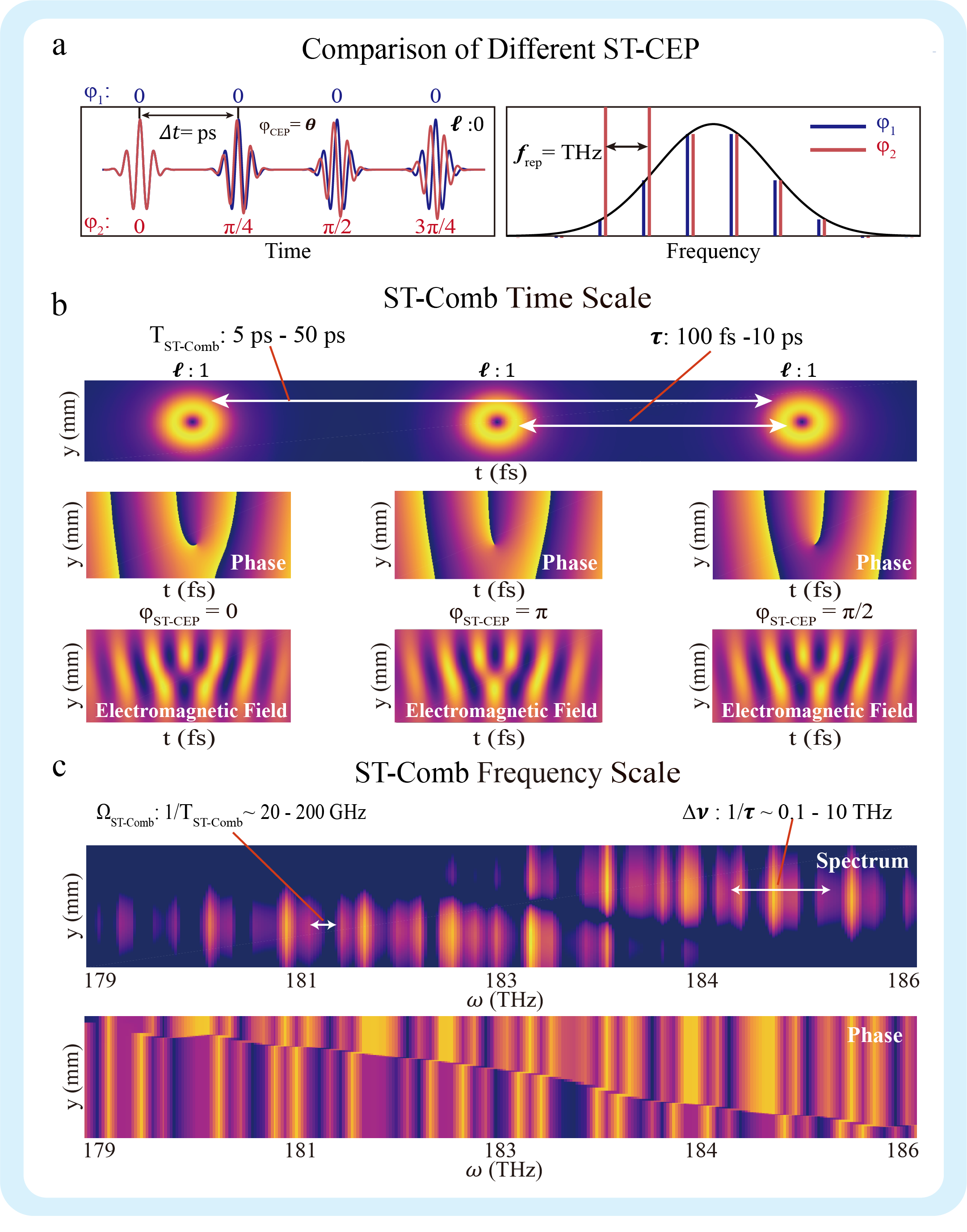}
\caption{\textbf{Temporal and Spectral properties of ST-CEP and ST-Comb.} \textbf{a,} Time- and frequency-domain waveforms under two ST-CEP settings. Tuning the ST-CEP enables THz-scale spectral sweeping. \textbf{b,} ST-Comb in the time domain. The comb forms a pulse train with $100$ fs–$10$ ps inter-pulse spacing and a total burst duration of $5$–$50$ ps, adjusting ST-CEP translates the electromagnetic-field envelope within the burst. \textbf{c,} ST-Comb in the frequency domain. The inter-pulse spacing $\tau$ sets the comb-line spacing in the sub-THz–THz band, yielding a dense comb, long-period modulation of the pulse train introduces an additional low-frequency spacing of ~$20$–$200$ GHz.}
 \label{Figure2}
\end{figure*}

Manipulating the carrier-envelope phase (CEP)—the phase between the carrier wave and the pulse's intensity envelope—enables sub-cycle waveform control, pushing light–matter interaction to its temporal limit and opening the door to petahertz-rate information processing \cite{krausz2014attosecond}. This work extends the concept of CEP beyond conventional Gaussian wavepacket temporal characteristics ($\psi_\text{CEP}$) to a spatiotemporal topological phase ($\psi_\text{ST-CEP}$),  within the ST-Comb framework. By modulating a single wavepacket with $\psi_\text{CEP}=\theta$ into a picosecond-spaced pulse train, each sub-pulse carries the same $\psi_\text{CEP}$. Meanwhile, the independently tunable phase $\psi_\text{ST-CEP}$ is encoded into each sub-pulse of the ST-Comb through a spatiotemporal modulation scheme (Figure~\ref{Figure2}a). Our approach controls the relative pulse waveform rather than the absolute electromagnetic field. It is critical to note that conventional CEP and the proposed spatiotemporal topological phase $\psi_\text{ST-CEP}$ coexist without mutual interference. While CEP imposes a global phase shift on the entire ST-Comb’s electromagnetic field, it does not alter the comb’s spectral intensity distribution. Because of this decoupling, $\psi_\text{ST-CEP}$ alone determines the ST-Comb spectrum, which is insensitive to CEP noise or lock state, making the method compatible with any femtosecond source without CEP stabilization.

In both the time and frequency domains, the ST-Comb displays a unique spatiotemporal topological structure. As shown in Figure~\ref{Figure2}b, its temporal structure is two-scale. A short scale ($100$ fs–$10$ ps) sets the inter-pulse delay $\tau$, and a long scale ($10$–$50$ ps) sets the total train duration. By Fourier duality, $\tau$ fixes the primary tooth spacing $ \Delta \nu = {1}/{\tau}$ in the sub-THz–THz range, while the finite train length introduces a secondary low-frequency spacing of tens of GHz. As an example, for an ST-Comb with topological number $[1; 1; 1]$, the field can be modeled as the convolution of a Gaussian pulse train with a spatiotemporal-topological wavepacket (Supplementary Materials 4), showing that the time-domain signal is jointly shaped by the Gaussian sequence and the ST-topological envelope. In the spectral domain (Figure~\ref{Figure2}c), the spectrum factorizes into the product of an optical frequency comb and a STOV spectrum. Fine comb lines arise from the long-period train, and the THz-scale envelope is fixed by $\tau$, yielding a composite spectrum spanning orders of magnitude.

\section*{Parameter-Space Encoding of High-Dimensional Information}
\begin{figure*}
\centering
\includegraphics[width=0.95\columnwidth]{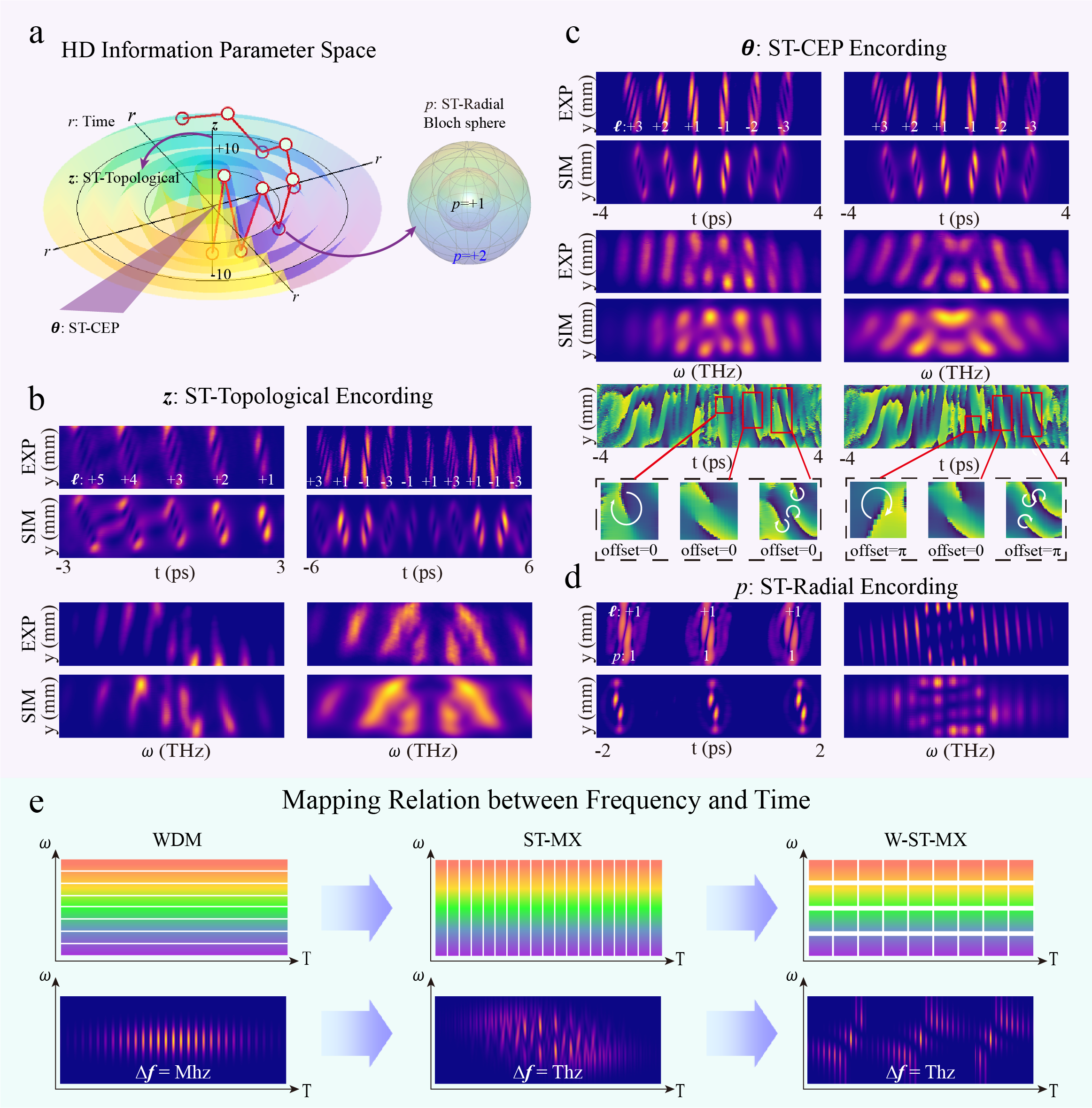}
\caption{\textbf{Degrees of freedom for high-dimensional manipulation of the ST-Comb and mapping relation from time–frequency resources.} \textbf{a,} Schematic of the high-dimensional information space of the ST-Comb. \textbf{b,} Spatiotemporal topological encoding (information-space $\mathit{z}$-axis). Examples showing ST-Combs with linear and nonlinear mappings between time and topological charge $\ell$. \textbf{c,} ST-CEP control (parameter-space $\theta$-axis). Varying the spatiotemporal carrier-envelope phase $\boldsymbol{\varphi}_{ST-CEP}$ preserves the time-domain waveform while inducing pronounced spectral differences. \textbf{d,}  Radial topological $\mathit{p}$ control. ST-Combs combining $\ell$ with the radial index $\mathit{p}$. \textbf{e,} Mapping Relation from frequency to time across three modes frequency multiplexing, spatiotemporal multiplexing, and hybrid frequency–spatiotemporal multiplexing.}
\label{Figure3}
\end{figure*}

To highlight the potential of the new degree of freedom, we construct a high-dimensional information parameter space using the ST-Comb (Figure~\ref{Figure3}a). The vertical axis $\mathit{z}$ encodes the spatiotemporal topological degree of freedom, the radius $\mathit{r}$ sets the inter-pulse spacing, and the polar angle $\theta$ represents the ST-CEP. Each point in this parameter space can be lifted along an additional dimension parameterized by the radial topological index $\mathit{p}$. Figure~\ref{Figure3}b–d present experimental and simulated control of each degree of freedom. Angular topology (Figure~\ref{Figure3}b), we show ST-Combs with different numbers of sub-pulses whose individual $\ell$ values either vary linearly in time or follow an aperiodic sequence, corresponding to coupled and uncoupled time–topological mappings (Supplementary Materials 5), which produce distinct temporal “chirality” signatures. By imparting a controllable spatiotemporal carrier-envelope phase (ST-CEP) through spectral encoding, this method generates ST-Combs that exhibit markedly distinct spectral fingerprints while preserving an identical time-domain structure due to their unchanged topology (Figure~\ref{Figure3}c, left vs right). By encoding information within a higher-dimensional topological state space, defined by variables such as the radial ($\mathit{p}$) and azimuthal ($\ell$) indices (Figure~\ref{Figure3}d), we elevate the structuring of spatiotemporal combs beyond the constraints of a single topological charge. This principle, demonstrated here with an orthogonal basis, fundamentally extends to non-orthogonal topologies \cite{pan2024non}, promising an even richer state space. This multi-dimensional control unlocks a vastly expanded parameter space (Supplementary Materials 6).

The coupling between time and frequency resources in physical systems is rooted in Fourier-transform duality. This duality enforces a dynamic constraint mechanism on the distribution of information across time and frequency domains, ensuring that the allocation of resources in one domain directly influences and limits the characteristics of the other, as quantitatively mapped by the Wigner distribution (Figure~\ref{Figure3}e). However, developing an optimal allocation framework that satisfies these physical constraints remains a critical theoretical challenge in advancing communication technologies\cite{kim2025complex}. Current information multiplexing schemes primarily focus on the frequency domain, such as WDM\cite{essiambre2010capacity}. The time domain has been increasingly overlooked, as traditional solitons lack programmable time structures, making it challenging to directly encode information in their temporal degrees of freedom. The ST-Comb enables discrete time-domain partitioning at fs-to-ps resolution, unlocking ultrafast temporal-structure modulation and high-dimensional information encoding via programmable spatiotemporal topologies.

\section*{Multiplexing and Demultiplexing of ST-Comb}
\begin{figure*}
\centering
\includegraphics[width=0.8\columnwidth]{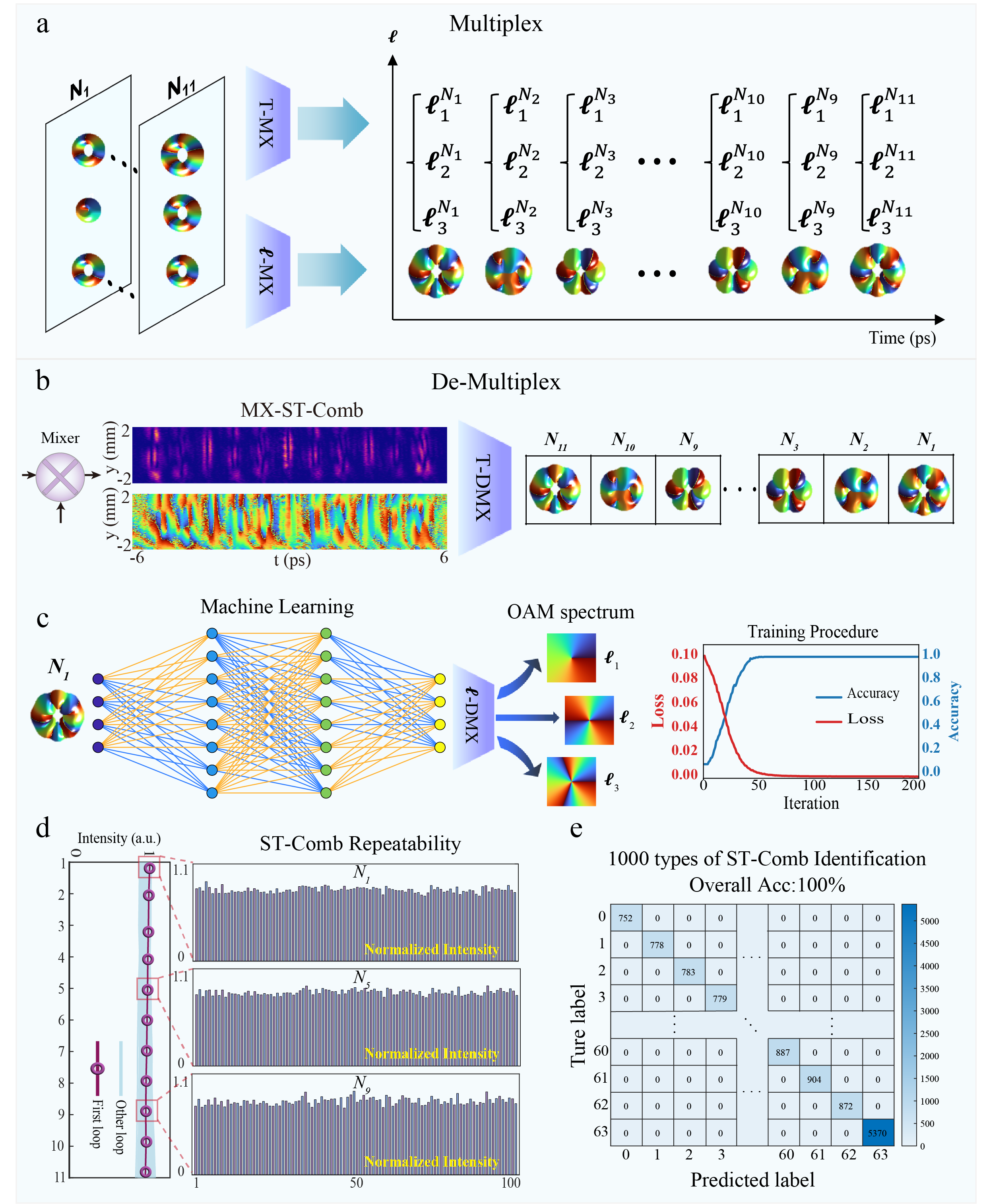}
\caption{\textbf{Multiplexing-Demultiplexing Mechanism of  ST-Comb.} \textbf{a,} Multiplexing process, femtosecond pulses undergo time multiplexing and spatiotemporal topological multiplexing to form hybrid-state ST-Comb. \textbf{b,} Time demultiplexing process, after coherent decoding hybrid-state ST-Combs undergo time demultiplexing to isolate individual sub-pulses. \textbf{c,} Topological demultiplexing process, time-demultiplexed sub-pulses are fed into machine learning for topological demultiplexing to recover original topological information. \textbf{d,} Repeatability demonstration, intensity stability results from $100$ repetitions of identical ST-Comb configurations \textbf{e,} Experimental measurement of $1000$ distinct ST-Combs achieves $100\%$ recognition accuracy.}
\label{Figure4}
\end{figure*}

Prior work shows that propagation can induce intrinsic splitting and mode conversion in spatiotemporal topology \cite{hancock2021mode}. This raises a key question, does such reshaping compromise the information encoded in ST-Comb? Here we show, theoretically and experimentally, that ST-Comb preserves a propagation-invariant topological quantity akin to a Winding number, so local rearrangements do not alter its global value. Rather than viewing the observed splitting of local topological charges as a limitation \cite{hancock2021mode}, we demonstrate that this behavior is governed by a more profound global principle. This fundamental invariance establishes the topological state as a robust unit for information, whose total identity persists through propagation. Building on this invariance, we construct an ST multiplexing-demultiplexing system that integrates phase-sensitive encoding and decoding with a deep neural network for high-fidelity recognition of complex topological states. The multiplexing process (Figure~\ref{Figure4}a) synergizes temporal multiplexing and ST-topological multiplexing, allowing each sub-pulse to carry topological superposition states. This extends the state capacity of the ST-Comb to $C = m n^{l}\quad (n \geq 2)$, where ${C}$ represents the total number of states, $\textit{m}$ denotes the number of ST-CEP states, $\textit{n}$ represents sub-pulses, and $\ell$ indicates topological charge. Relative phase differences between superposed states could also provide an additional encoding degree of freedom (Supplementary Materials 7). Experimentally, we demonstrated parallel transmission of multiple independent, crosstalk-free sub-pulses in the temporal domain (Supplementary Materials 8). For topological states, we showed superpositions spanning a wide range of charges and their combinatorial configurations, enabling a single dimension to encode an exponentially large number of distinct states. By coherently controlling both temporal and topological dimensions, we constructed a vast state space theoretically supporting ultrahigh data capacities. Current scaling limitations arise primarily from constraints in CCD photosensitive areas, dynamic range, and spatial light modulator pixel resolution.

The high-dimensional state space in the ST-Comb framework—capable of hosting billions of unique states—is analogous to the uniqueness of human fingerprints, where each state exhibits distinct identifiability. However, the near-impossibility of direct decoding within this vast space necessitates picosecond-resolution temporal slicing of complete pulses to achieve accurate state identification. Leveraging the ST-Comb’s time-separated, crosstalk-free sub-pulse characteristics, we employed time-division demultiplexing to reduce computational complexity (Figure~\ref{Figure4}b). This approach reduces the effective complexity of decoding from exponential growth to linear scaling. Figure~\ref{Figure4}c summarizes the machine learning training process and feature extraction framework for post-decoding ST-Comb analysis. The method achieved near-unity classification accuracy across randomly generated ST-Comb samples (Figure~\ref{Figure4}e), demonstrating strong decoding fidelity. Robustness was further validated by consistent intensity stability across repeated trials (Figure~\ref{Figure4}d).

\section*{Discussion}
In this work, we have developed a spatiotemporal topological optical frequency comb framework that provides a unified route to high-dimensional information encoding. Drawing on sonar\cite{misaridis2005use} and radar\cite{woodward2014probability} paradigms, the approach encodes signals at THz–PHz optical carriers and demodulates them within a spectrally cleaner region, reducing the impact of low-frequency noise. Through All-DoF modulation, spatiotemporal topological degrees of freedom can be independently tuned, creating opportunities for dense encoding strategies. Importantly, the framework still contains unexplored degrees of freedom: sub-pulse intensities, higher-order radial indices, and relative phase relations between superposed states. These collectively generate a combinatorial state space with scaling properties that significantly expand the encoding potential, positioning ST-Comb as a candidate platform for multi-dimensional “optical fingerprints” in information labeling and storage. Beyond communications, this capability connects to fundamental science, including sub-cycle control in PHz electronics\cite{krausz2014attosecond}, explorations of high-dimensional nonlinear physics\cite{wright2022physics}, and possible extensions of programmable ST-Combs into the extreme-ultraviolet (EUV)\cite{gohle2005frequency}.

From a practical standpoint, the ST-Comb concept is compatible with ongoing advances in integrated photonics. It naturally aligns with on-chip frequency combs\cite{trocha2018ultrafast,zhao2024all}, ultrafast modulators\cite{li2019phase,panuski2022full}, programmable active metasurfaces\cite{shaltout2019spatiotemporal}, and optical phased arrays\cite{morimoto2018diffraction,hu2024hyper}. Together, these technologies provide a feasible hardware basis for implementing energy-efficient, high-capacity photonic systems (Supplementary Material 12). More broadly, full spatiotemporal control may stimulate progress in optical computing architectures\cite{xu2022multichannel}, photonic artificial intelligence \cite{xu2024large,xue2024fully}, and reconfigurable high-dimensional light-field processors. Taken together, these directions highlight a path from laboratory-scale demonstrations toward practical, chip-scale platforms for future reconfigurable photonics.

\section*{Method}
\subsection*{Experimental Setup}
All experiments were performed using a Yb-based femtosecond laser (central wavelength $1030$ $\mathrm{nm}$; spectral FWHM $\approx 20$ $\mathrm{nm}$; transform-limited pulse duration $\approx 170$ $\mathrm{fs}$). The output beam was divided into a signal arm and a probe arm by a dielectric beam splitter. In the signal arm, a folded two-dimensional spatiotemporal pulse modulator was used to imprint user-defined complex amplitude and phase profiles onto the pulse. The modulator consisted of three cascaded elements: (i) a reflective diffraction grating ($1200$ $\mathrm{lines/mm}$) to angularly disperse the broadband spectrum. (ii) a cylindrical lens ($f = 200$ $\mathrm{mm}$, axis parallel to the grating grooves) to focus the dispersed spectrum along the \textit{x}-axis, and (iii) a phase-only reflective spatial light modulator (Holoeye GAEA) positioned at the Fourier plane. A precalculated complex transmittance pattern was uploaded to the SLM, producing a spatiotemporal optical frequency comb in which each sub-pulse carried a programmable topological charge.

\subsection*{Mathematical formulation of the ST-Comb}
A spatiotemporal optical vortex comb (ST-Comb) can be synthesized by coherently superposing optical-field degrees of freedom (DoF) in both space and time. This process yields a train of structured wave packets with tightly coupled spatiotemporal vortex properties. Such packets are generally nonseparable, meaning the field cannot be expressed as a simple product of a spatial mode $L(\mathbf{r})$ and a temporal envelope $T(t)$. The electric field can be written as
\begin{equation}
E(x, y, t) = \sum_{n} C_n\, A_n(x, y; \mu_n) \exp \left[ i \big( \omega_0 t + \psi_{\mathrm{ST\text{-}CEP}}(n) \big) \right],
\end{equation}
where $A_n(x, y; \mu_n)$ denotes the $n$-th spatiotemporal mode, $C_n$ is the complex weighting coefficient, $\omega_0$ is the carrier frequency, and $\psi_{\mathrm{ST\text{-}CEP}}(n)$ is the spatiotemporal carrier–envelope phase of the $n$-th sub-pulse.

The ST-Comb can also be modeled as the convolution of a complex spatial amplitude distribution with a periodic Dirac comb along the time axis, capturing its spatiotemporal periodicity and vortex structure. Conservation of angular momentum under mapping from the $(k_x,\omega)$ to $(x,t)$ domains ensures that the topological phase remains invariant. By projecting onto the angular-momentum and temporal axes, an $\ell$–$t$ distribution is obtained, which reveals how programmable coefficients $C_n$ control both sub-pulse amplitudes and topological charge. This representation highlights the high-dimensional $\ell$–$t$ manifold available for encoding, demonstrating the scalability of the ST-Comb framework.

\bibliographystyle{unsrt}
\bibliography{ref(new)}

\subsection*{Acknowledgments}
This work was supported by National Natural Science Foundation of China (12374318 [G.F.], 12434012 [Q.Z.], 12450407 [Z.T.], 12274091 [Z.T.], 12474336 [Q.C.]), National Key Research and Development Program of China (Grant Nos. 2021YFA1400200 [Z.T.]), Science and Technology Commission of Shanghai Municipality (22JC1400200 [Z.T.], 24JD1402600 [Q.Z.], 24QA2705800 [Q.C.]), Key Project of Westlake Institute for Optoelectronics (Grant No. 2023GD007 [Q.Z.]). We acknowledge the Shanghai Astronomical Observatory, Chinese Academy of Sciences, for providing access to their experimental facilities during part of this work.
\subsection*{Author contributions} 
D.L. and G.F. proposed the original idea and initiated this project. D.L. completed the theory and simulations. D.L. designed and performed the experiments. H.W. and D.L. both analyzed the data. D.L., H.W., G.F., and Q.Z. prepared the manuscript. All authors contributed to the discussion and writing of the manuscript.

\subsection*{Competing interests}
The authors declare no competing interests.

\subsection*{Data availability}
The data that support the findings of this study are available from the corresponding authors on reasonable request.

\subsection*{Additional information}
\textbf{Correspondence and requests for materials} should be addressed to Guangyu Fan.

\baselineskip 21pt
\clearpage

\end{document}